\documentclass[12pt]{iopart}

\begin{document}

\title[The kinematical Hilbert space of Loop Quantum Gravity]{The kinematical Hilbert space of Loop Quantum Gravity from BF theories}

\author{Francesco Cianfrani}

\address{Dipartimento di Fisica, Universit\`a  di Roma ``Sapienza'', Piazzale Aldo Moro 5, 00185 Roma, Italy.}
\ead{francesco.cianfrani@icra.it}
\begin{abstract}
In this work, it is demonstrated how the kinematical Hilbert space of Loop Quantum Gravity (LQG) can be inferred from the configuration space of BF theories via the imposition of the Hamiltonian constraints. In particular, it is outlined how the projection to the representations associated with Ashtekar-Barbero connections provides the correct procedure to implement second-class constraints and the corresponding nontrivial induced symplectic structure. Then, the reduction to SU(2) invariant intertwiners is analyzed and the properties of LQG states under Lorentz transformations is discussed. 
\end{abstract}

\pacs{04.60.Pp}

\maketitle

\section{Introduction}

Spin-foam models \cite{rov} are based on the idea that 4-dimensional gravity behaves as a topological BF theory with some constraints \cite{bfsf}. In particular, the theory is discretized on a simplicial complex and then quantized, thanks to discretization independence \cite{cy}. BF theories are based on a covariant formulation, in which basic configuration variables belong to the SL(2,C) algebra. The SU(2) gauge structure proper of Loop Quantum Gravity (LQG) comes out via the imposition of the constraints which reduce the BF action to the Holst one. These additional conditions belong to a second-class set, whose treatment on a quantum level is not free of ambiguities and leads to different scenarios. The most relevant cases are the Engle-Livine-Pereira-Rovelli (EPRL) \cite{eprl} and the Freidel-Krasnov (FK) \cite{fk} models.   

The intermediate step between the Lorentz covariant formulation of BF theories and LQG is realized through projected spin-networks \cite{psn}, which provide a map from SU(2) invariant states to SL(2,C) functions. Such a representation of SU(2) spin-network in terms of SL(2,C) functions gives the tools to demonstrate the Lorentz invariance of the whole formulation \cite{recrov} (see also \cite{wv}). Projected spin-networks are not normalizable in the SL(2,C) scalar product, but they are so in the SU(2) one. This is not surprising, in view of the presence of constraints.

However, all the known procedures to impose second-class conditions on a quantum level are questioned in \cite{alsf}. In this work, Lorentz irreducible representations of the principal series are restricted to a certain subset, associated with Ashtekar-Barbero connections, such that the simplicial constraints are solved. The criticism is based on the idea that this procedure to solve second-class constraints on a quantum level is inconsistent with a reduced phase space quantization, because it does not account for the nontrivial induced symplectic structure. 

In this work, we discuss a procedure to define the LQG kinematical Hilbert space from the configuration space of BF theories in a continuum setting. This framework is based on LQG without the time gauge \cite{prl} and it allows us to demonstrate that the issues discussed in \cite{alsf} do not apply to gravity. In particular, we review the Hamiltonian formulations of BF and Holst actions and we outline how the proposal made by Alexandrov in \cite{alsf} to infer the Ashtekar-Barbero connections provides the same reduction given classically by the explicit solution of second-class constraints. In particular, the modification of the induced symplectic structure due to the second-class character of constraints is reproduced at the level of quantum configuration variables. Hence, the projection on a proper subspace inside each Lorentz irreducible representation is able to account for the presence of second-class constraints on a quantum level and no inconsistency arises. 

Then, we depict how the quantization can be carried on in this scheme by imposing the Hamiltonian constraints according with the Dirac procedure. The distinctive feature of this model, with respect to the one discussed in \cite{recrov}, is that there exists a preferred embedding of the Lorentz group in the SU(2) one. In fact, the group of rotations on boundary (spatial) hypersurfaces constitutes a privileged SU(2) subgroup because the associated connections contain all dynamical information on the gravitational system. In this respect, we develop quantum states as the tensor product of SU(2) holonomies at edges and functionals of the boost parameters at vertices. The implementation of the vanishing behavior for the Hamiltonian constraints fixes the dependence from boost parameters and provides the reduction of the intertwiners to the SU(2) invariant ones. This feature completes the derivation of the kinematical Hilbert space proper of LQG from BF theories, such that the proposed scheme constitutes an alternative description with respect to the one given by projected spin-networks and implemented in Spin-Foam models. Finally, the behavior of rotation and boost operators is discussed and the invariance of physical states under their action is outlined. 

Therefore, a consistent correspondence is established between LQG and BF theory at a kinematical level, while the dynamical features of these two models still need to be discussed properly in the adopted scheme. This issue will require the extension of this work to a path-integral formulation. 

In the following, latin capital letters $I, J, K$ and latin lowercase letters $i, j, k$ will denote Lie algebra indexes for the Lorentz group and the SU(2) group, respectively, while $\mu, \nu, \rho$ and $a, b, c$ will denote tensorial space-time and spatial indexes, respectively.

\section{BF theories vs LQG without the time gauge}

In the BF theory \cite{bf} the configuration variables are the 2-form $B^{IJ}$ and the Lorentz connection 1-form $\omega^{IJ}$ and the action is given by 
\begin{equation}
S=\frac{1}{2}\int \epsilon^{\mu\nu\rho\sigma} tr(B^{IJ}_{\mu\nu}F^{KL}_{\rho\sigma})d^4x,
\end{equation}

$F^{KL}$ being the curvature of $\omega^{IJ}$, while the trace is realized by contracting internal indexes with the invariant metric of the Lorentz group $g_{IJKL}=\eta_{I[K}\eta_{L]J}$. The 2-form $B^{IJ}$ enters the definition of the conjugate momenta to the components $\omega^{IJ}_a$ as follows 
\begin{equation}
\Pi^a_{IJ}=g_{IJKL} B^{KL}_{bc}\epsilon^{abc},
\end{equation}

and the full Hamiltonian reads
\begin{equation}
H=\int \left[-\frac{1}{2}\epsilon^{abc}tr(B^{IJ}_{tc}F^{KL}_{ab})-\omega^{IJ}_tD_a\Pi^a_{IJ}\right]d^3x,\label{ham}
\end{equation}

where $D_a$ denotes the covariant derivatives built from $\omega^{IJ}_a$.

Because $B^{IJ}_{tc}$ and $\omega^{IJ}_t$ are independent variables with no dynamical role, the Hamiltonian is a linear combination of the constraints 
\begin{equation}
\left\{\begin{array}{c} D_a\Pi^a_{IJ}=0 \\ F^{KL}_{ab}=0  
\end{array}\right.,
\end{equation}

where the first condition implements the local Lorentz invariance, while the other one ensures the topological character of the theory. They realize a first-class set of constraints.

The Holst action can be written formally as a BF theory, but the 2-form $B^{IJ}$ is developed from the 1-form $e^I$ of the space-time metric $g_{\mu\nu}$ in the following way
\begin{equation}
B^{IJ}=\sqrt{g}{}^\gamma\!p^{IJ}_{KL} \ast(e^K\wedge e^L)= \sqrt{g}\left[\delta^{IJ}_{KL}-\frac{1}{2\gamma}\epsilon^{IJ}_{\phantom{12}KL} \right]
\ast(e^K\wedge e^L), \label{Be}
\end{equation}

$\gamma$ being the Immirzi parameter. The relation (\ref{Be}) implies that the momenta $\Pi^a_{IJ}$ are not independent and 
as soon as one defines
\begin{equation}
\Pi^a_{IJ}=\left[\delta_{IJ}^{KL}-\frac{1}{2\gamma}\epsilon^{KL}_{\phantom{12}IJ} \right] \pi^a_{KL}, 
\end{equation}

the following conditions hold 
\begin{equation}
C^{ab}=\epsilon^{IJKL}\pi^a_{IJ}\pi^b_{KL}=0.
\end{equation}

As a consequence of the relation (\ref{Be}), $B^{IJ}_{tc}$ are not independent anymore and the term $-\frac{1}{2}\epsilon^{abc}tr(B^{IJ}_{tc}F^{KL}_{ab})$ can be written as 
\begin{equation}
\frac{1}{g^{tt}}H+\frac{g^{ta}}{g^{tt}}H_a,
\end{equation}

$H$ and $H_a$ being in a 3+1 representation the super-Hamiltonian and the super-momentum, respectively, whose expressions read
\begin{equation}
H=\pi^a_{IK}g^{KL}\pi^{b}_{LJ}{}^\gamma\!p^{IJ}_{\phantom1\phantom2MN}F^{MN}_{ab},\qquad H_a={}^\gamma\!p^{IJ}_{\phantom{12}KL}\pi^b_{IJ}F^{KL}_{ab}.
\end{equation} 

Additional conditions are provided by the nonvanishing behavior of the Poisson brackets between $C^{ab}$ and $H$, {\it i.e.}
\begin{equation}
D^{ab}=[C^{ab},H]=\epsilon^{IJKL}g^{MN}\pi^c_{IM}\pi^{(a}_{NJ}D_k\pi^{b)}_{KL}, 
\end{equation}

such that the set of constraints is given by \cite{lr}
\begin{equation}
\left\{\begin{array}{c} D_a\Pi^a_{IJ}=0 \\ H=0 \\ H_a=0 \\ C^{ab}=0 \\ D^{ab}=0
\end{array}\right.,
\end{equation}

and it is second class, because $[C^{ab},D^{ab}]$ and $[D^{ab}, D^{ab}]$ do not vanish on-shell. As a consequence, the induced symplectic structure on the hypersurfaces where $C^{ab}=D^{cd}=0$ can be non-trivial.

The possibility to reduce a second-class set of constraints to a first-class one by a proper choice of phase-space variables constitutes the main property of such a kind of conditions. In fact, the constraints which make a system second-class are not related with gauge symmetries, but they just signal that a choice of coordinates exists in which some variables become redundant. A different interpretation for second-class constraints is that they arise in the presence of gauge fixings. A system of second-class constraints can be described by solving the constraints themselves and working with reduced variables or by using the full set of coordinates and replacing Poisson brackets with Dirac ones.   

In \cite{prl}, the former procedure has been adopted and it has been demonstrated that it is possible to develop a formulation in terms of first-class constraints only where 
\begin{itemize}
{\item no gauge fixing of the local Lorentz frame takes place,}
{\item the Gauss constraints of the Lorentz group is mapped into the SU(2) Gauss constraints associated to Ashtekar-Barbero connections $A^i_a$ and densitized triads $E^a_i$, while additional conditions are provided by the vanishing of the conjugate momenta $\pi^i$ to boost parameters $\chi_i$, {\it i.e.}
\begin{equation}
G_i=\partial_aE^a_i+\gamma\epsilon_{ij}^{\phantom{12}k}A^j_aE^a_k=0,\qquad \pi^i=0.\label{lsp}
\end{equation}
}
\end{itemize}

In particular, the solution to second-class constraints via a Lorentz transformation, which formally restores the time gauge, can be written as
\begin{equation}
\left\{\begin{array}{c}\pi^a_{ij}=0,\\ \omega^{ij}_{a}={}^E\!\omega^{ij}_{a}
\end{array}\right.,
\label{scon}
\end{equation}

${}^E\!\omega^{ij}_{a}$ being the spin connections associated to $E^a_i$. While in the whole phase-space the induced symplectic structure is trivial,  once the restriction to the hypersurfaces (\ref{scon}) takes place the following Poisson brackets are induced
\begin{eqnarray}
\{E_{i}^a(x,t),E_{j}^a(y,t)\}=0\label{1c}\\
\{\omega^{0i}_a(x,t),\omega^{0j}_b(y,t)\}=-\frac{1}{\gamma}\delta^{[i}_{k}\epsilon^{j]}_{\phantom1lm}\frac{\partial{}^E\!\omega^{lm}_j(y,t)}{\partial E^a_k(x,t)}\label{2c}\\
\{\omega^{0i}_a(x,t),\pi_j^b(y,t)\}=\delta^b_a\delta^3(x-y)\delta^{i}_{j}.\label{3c}
\end{eqnarray}
 
These modified Poisson brackets can be reduced to the trivial ones by choosing as coordinates $A^i_a=\omega^{0i}_a+\frac{1}{2\gamma}\epsilon^i_{\phantom{12}jk}{}^E\!\omega^{jk}_a(E)$ and $E^a_i$, {\it i.e.}  
\begin{eqnarray}
\{ A^i_a(x,t),E^b_j(y,t) \} =\delta^i_j\delta^b_a\delta^3(x-y),\label{ss1}\\ \{ A^i_a(x,t),A^j_b(y,t) \} = \{ E^a_i(x,t),E^b_j(y,t) \} =0.\label{ss2}
\end{eqnarray}

In other words, it is possible to get rid of second-class constraints by restricting the full phase space to the hypersurfaces where the conditions (\ref{scon}) hold and by adopting the symplectic structure given by the relations (\ref{ss1}) and (\ref{ss2}). Because the conjugate variables associated with the boost parameters vanish, the full kinematical phase-space can be described by the coordinates $A^i_a,E^a_i$ only. The resulting quantum theory is developed from the space of cylindrical functionals associated to the SU(2) holonomies along graphs and the full Hilbert space at each node can be decomposed as the direct product of SU(2) irreducible representations. 

Given a surface $S$ and an edge $e$, whose initial point belongs to $S$, the momentum operator smeared on $S$ acts on the holonomy along $e$ as follows (in units $\hbar=c=8\pi G=1$) 
\begin{equation}
E_i(S)h^{(j)}_{e}=-i\gamma h^{(j)}_{e}\tau^{(j)}_i o(S,e),\label{su2}
\end{equation}

where $o(S,e)$ is the factor equal to $1,-1, 0$ if the normal to S and the tangent to $e$ are collinear, anticollinear or coincident, respectively, while $\tau^{(j)}_i$ denote the SU(2) generators with spin $j$. The equation above is the quantum counterpart of the conditions (\ref{ss1}), thus once implemented on a quantum level it is able to account properly for the restriction to the hypersurfaces in which second-class constraints hold. 

\section{Second-class constraints on a quantum level}

Here we depict how in the recent formulation of spin-foam models (\cite{alsf}) the restriction to the hypersurfaces (\ref{scon}) takes place on a quantum level, linking BF theory with LQG. 
Configuration variables associated with BF theory in SF models are defined in analogy with LQG 
. Hence, the configuration space is the one of cylindrical functionals over graphs, whose building blocks are holonomies along edges. Because holonomies transform only at the initial and final points of the edges, the Hilbert space at each node is developed as the tensor product of one Lorentz representations for each edge and it is equipped with the Haar measure associated to the universal covering of SO(1,3). By the Peter-Weyl theorem the Hilbert space can be decomposed as the direct integral over irreducible representations. The gauge group being non-compact, the definition of an invariant scalar product cannot be accomplished by standard techniques and the boundary states are usually represented by non normalizable functions, from which the kinematical Hilbert space of LQG is inferred by a proper reduction (see \cite{recrov} and references therein).  
The irreducible representations of the Lorentz group belonging to the principal series are labeled by an half-integer number $k$ and a real one $\rho$ and they can be split in terms of SU(2) ones as follows
\begin{equation}
H^{(k,\rho)}=\oplus_{j=k}^{+\infty} H^{(j)}.
\end{equation}  

The action of the operators associated to rotation $R_i=\epsilon_i^{\phantom1jk}T_{jk}$ and boosts $K_i=T_{0i}$ on each $H^{(j)}$ are given by \cite{lr}
\begin{equation}
\left\{\begin{array}{c}
R_+|j,m>=\sqrt{(j-m)(j+m+1)}|j,m+1>, \\
R_-|j,m>=\sqrt{(j+m)(j-m+1)}|j,m-1>, \\ 
R_3|j,m>=m|j,m>, \\
K_+|j,m>=-\alpha_j\sqrt{(j-m)(j-m-1)}|j-1,m+1>-\\-\beta_{j}\sqrt{(j-m)(j+m+1)}|j,m+1>-\alpha_{j+1}\sqrt{(j+m+1)(j+m+2)}|j+1,m+1>,\\
K_-|j,m>=\alpha_j\sqrt{(j+m)(j+m-1)}|j-1,m-1>-\\-\beta_{j}\sqrt{(j+m)(j-m+1)}|j,m-1>+\alpha_{j+1}\sqrt{(j-m+1)(j-m+2)}|j+1,m-1>,\\
K_3|j,m>=-\alpha_j\sqrt{j^2-m^2}|j-1,m>-\\-\beta_{j}m|j,m>+\alpha_{j+1}\sqrt{(j+1)^2-m^2}|j+1,m>,\\
\end{array}\right.
\end{equation}

where 
\begin{equation}
\alpha_{j}=\frac{i}{j}\sqrt{\frac{(j^2-k^2)(j^2+\rho^2)}{4j^2-1}},\qquad \beta_{j}=\frac{k\rho}{j(j+1)}.
\end{equation}

In BF theory the symplectic structure in the whole phase-space is trivial, thus the action of the operator associated to smeared momenta is found by promoting the Poisson brackets (and not the Dirac ones) to commutators. In particular one finds
\begin{equation}
\Pi_{IJ}(S)h^{(\rho,k)}_{e}=i h^{(\rho,k)}_{e}T^{(\rho,k)}_{IJ} o(S,e),\label{Pq}
\end{equation}

$T^{(\rho,k)}_{IJ}$ being the generators of the Lorentz group in the representation $(\rho,k)$.

Alexandrov \cite{alsf} noted that it is possible to project down Lorentz irreducible representations to the ones associated with the Ashtekar-Barbero SU(2) connections via the projector $\pi^{(j)}$ to the spin $j$ part when 
\begin{equation}    
	\beta=\frac{k\rho}{j(j+1)}=\gamma.\label{con}
\end{equation}

This condition fixes the spin number of the selected SU(2) representation inside each Lorentz one, such that the projector reads
\begin{equation}
\pi^{(j)}: h^{(\rho,k)}\rightarrow {}^L\!h^j=h^{(\gamma \frac{j(j+1)}{j-r},j-r)}.\label{rep}
\end{equation}

In the expression above $r$ denotes the nonnegative half-integer parameter such that $j=k+r$. Hence, given a spin $j$ SU(2) representation, it can be inferred from all Lorentz irreps $\{\rho,k\}$ for which a positive half-integer parameter $r$ exists such that $\rho=\gamma(j(j+1))/(j-r)$ and $k=j-r$. In other words, there are many Lorentz irreps associated to a given SU(2) one and they are labeled by $r$. This redundancy is due to the fact that there is only one condition (\ref{con}) for the two parameter $\{\rho,k\}$.

By applying the operator associated to the smeared momenta on $h_L^{(j)}$ (\ref{Pq}), the following conditions are found 
\begin{eqnarray}
\pi_{ij}(S){}^L\!h^{(j)}_e=0,\\ 
\pi_{0i}(S){}^L\!h^{(j)}_e=-i\gamma {}^L\!h^{(j)}_{e} \tau_i o(S,e),
\end{eqnarray}
  
It is worth noting that 
\begin{itemize}
{\item the holonomies ${}^L\!h^{(j)}$ are in the kernel of the operator associated to the first condition in the system (\ref{scon});}    
{\item the action of $\pi_{0i}(S)=E_i(S)$, coincides with the one obtained when $h^{(j)}$ are holonomies of the Ashtekar-Barbero connection (\ref{su2}).}
\end{itemize}


This analysis demonstrates that the condition (\ref{con}) provides not only the solutions of second-class constraints, but also that the action of flux operators coincide with the one of LQG. Because such an action is determined by the symplectic structure in phase-space, this procedure to solve constraints is able to account for the nontrivial induced symplectic structure on constraint hypersurfaces.  In other words, the restriction to the irreps (\ref{rep}) implements properly the features of second-class constraints (reduction to a proper phase-space hypersurfaces and nontrivial induced symplectic structure), such that the objection to the whole quantization procedure raised in \cite{alsf} does not hold. Therefore, the findings of this work support the viability of the spin-foam models as the proper quantum description of the gravitational field.     

It is worth noting that by fixing the spin quantum number $j$, there is a degeneracy given by the parameter $r$ inside (\ref{rep}). The resulting space-time structure has been investigated in \cite{gen}. The additional restriction to $r=0$ does not provide any modification on the statements above about the consistency of the procedure to solve constraints.   

\section{The kinematical Hilbert space of LQG}

Within this scheme, a fundamental dependence from boost parameters remains at vertices. Let us consider a 3-valent vertex, which according to recoupling theory constitutes the building block of any $n$-valent vertex. The intertwiner sending $|\rho_1,k_1, j_1,m_1>\otimes |\rho_2,k_2, j_2,m_2>$ to $|\rho,k,j,M>$ coincide with the Clebsch-Gordan coefficient of the Lorentz group, which can be written as follows \cite{cbl}
\begin{equation}
i_{\rho_1,k_1, j_1,m_1,\rho_2,k_2, j_2,m_2}^{\rho,k,j,M}=
(N^{\rho k \rho_1 k_1 \rho_2 k_2}_{j'M' j_1'm_1' j_2'm_2'})^{-1} 
\int d\mu(g) D^{\rho k *}_{jM, j'M'}(g)D^{\rho_1 k_1}_{j_1m_1, j_1'm'_1}(g)D^{\rho_2 k_2}_{j_2m_2, j'_2m'_2}(g),\label{CB}
\end{equation} 

$N^{\rho k \rho_1 k_1 \rho_2 k_2}_{j'M' j_1'm_1' j_2'm_2'}$ being a normalization term, while the integration is extended over the whole Lorentz group with the Haar measure $d\mu(g)$ and $D^{\rho k}_{jm, j'm'}(g)$ denotes the irreducible representations. Each element of the Lorentz group can always by written as a rotation times a boost and the irreducible representations can be split according with such a decomposition, {\it i.e.}
\begin{equation}
D^{\rho k}_{jm, j'm'}(g)=\sum_{\lambda=-min(j,j')}^{min(j,j')} D^{\rho k}_{jj'\lambda}(\chi)D^{j'}_{\lambda m'}(\phi_1,\theta,\phi_2).
\end{equation}

The integration over the group manifold inside the expression (\ref{CB}) can be split into the one over rotation parameters 
and the one over boost parameters $\chi_i$.
Hence, to each boundary point of edges is attached the product of boost representations and the full vertex is constructed from these basic states in such a way that the final intertwiner coincides with the Lorentz one (\ref{CB}). The presence of the variables $\chi_i$ at each vertex is similar to the case of projected spin networks \cite{psn}. However, having solved second-class constraints, the edges carry SU(2) representations and not Lorentz ones, while the boost parameters merely enter the definition of the intertwiners.
In order to define Lorentz-invariant states, let us consider the representation of the Lorentz group acting on the variables $\{A^i_a,\chi_i\}$ \cite{prl}, {\it i.e.}
\begin{eqnarray} 
R_i=G_i+\epsilon_{i\phantom{1}k}^{\phantom{1}j}\chi_j\pi^k,\label{rot}\\ K_i=(\delta_i^j+\chi_i\chi_j)\pi^j-\beta\epsilon_i^{\phantom{1}jk}\chi_jG_k \label{boo}
\end{eqnarray}

$G_i$ being the Gauss constraint of the SU(2) group, while $\beta=\frac{1+\sqrt{1-\chi^2}}{\chi^2}$\footnote{Here $\chi^2=\sum_i \chi_i\chi_i>0$, so we change notation with respect to \cite{prl}}.

Because $A^i_a,\chi_i$ commute, a generic state defined on a graph $\alpha$ can be represented as 
\begin{equation}
\psi_\alpha(A^i_a,\chi_i)=\otimes_e {}^L\!h_e(A^i_a) \otimes_v I_v(\chi_i),\label{st}
\end{equation}

where $e$ and $v$ denote the edges and vertices of $\alpha$, respectively. The new feature of this scenario is that $A^i_a$ are precisely the SU(2) Ashtekar-Barbero connections. Hence, the covariant description implies the enlargement of the configuration space to the variables $\chi_i$ describing the boost parameters of the local frame. 

On a classical level, $\chi_i$ are coordinates on a hyperbolic space, whose associated metric tensor is given by $\delta_{ij}+\frac{\chi_i\chi_j}{1-\chi^2}$. 

This way, the Hilbert space can be defined as the direct product of the one of distributional connections proper of LQG times the space of square-integrable functions $I_v(\chi)$ defined on the hyperbolic space parametrized by $\chi_i$. This space can be equipped with the following scalar product
\begin{equation}
<I^1_v(\chi),I^2_v(\chi)>=\int \frac{1}{\sqrt{1-\chi^2}} I^{1*}_vI^2_v d\chi_1d\chi_2d\chi_3,\label{scl}
\end{equation}

where the factor $1/\sqrt{1-\chi^2}$ is the determinant of the metric tensor. With such a scalar product the symmetric operators associated with momenta reads as follows
\begin{equation}
\pi^i=-i\left(\frac{\partial}{\partial\chi_i}+\frac{\chi_i}{2(1-\chi^2)}\right).
\end{equation}

In fact, given such a definition one can verify that $<I^1_v,\pi^iI^2_v>=<\pi^iI^1_v,I^2_v>$.

This way, the scalar product in the full Hilbert space reads as
\begin{eqnarray}
<\psi^1_\alpha,\psi^2_\alpha>=\otimes_{e}\int {}^L\!h^{1\dag}_e{}^L\!h^2_e d\mu_{SU(2)} \int \frac{1}{\sqrt{1-\chi^2}} I^{1*}_vI^2_v d^3\chi,
\label{scpr}
\end{eqnarray}

where $d\mu_{SU(2)}$ denotes the Haar measure associated with the SU(2) group.

\subsection{LQG states}

The states associated with LQG can be inferred by applying the constraints (\ref{lsp}) to the functions (\ref{st}). In particular the vanishing behavior of conjugate momenta to boost parameters gives 
\begin{equation}
\pi^i I^{LQG}_v=0\rightarrow I^{LQG}_v(\chi)\propto(1-\chi^2)^{1/4}, \label{pi}
\end{equation}

and if this solution is inserted into the expression (\ref{scpr}), the factor $1/\sqrt{1-\chi^2}$ is eliminated and the scalar product proper of LQG is inferred.

The condition $G_i=0$ coincides with the Gauss constraint proper of SU(2) gauge theory, thus it can be implemented in the space of distributional connections by inserting invariant intertwiners at vertices. 

Therefore, it is possible to define LQG states replacing $I_v(\chi)$ into the expression (\ref{st}) with SU(2) invariant intertwiners and neglecting any dependence on $\chi_i$ variables at vertices. This analysis completes the derivation of the kinematical Hilbert space proper of LQG in a covariant setting.

\subsection{The rotation generator}

The properties of $\psi_\alpha(A^i_a,\chi_i)$ under rotations can be inferred by noting that the generators $R_i$ (\ref{rot}) are the sum of the SU(2) generators associated with Ashtekar-Barbero connections and the orbital angular momenta of $\chi_i$. Furthermore, $R_i$ act at vertices only. 

In order to label states according with their properties under rotations, let us now consider a single vertex $v$ and the set $e(v)$ of edges incoming in it (for simplicity let us assume that the edges are all incoming). The holonomy associated to each $e(v)$ can be expanded in irreducible SU(2) representations as follows
\begin{equation}
\otimes_{e(v)} {}^L\!h_{e(v)}(A^i_a)=\otimes_{e(v)} \sum_{j_{e(v)}} c(j_{e(v)}) {}^L\!h^{j_{e(v)}}_{e(v)}
\end{equation}

and each term of this sum transforms under rotations according with the $j_{e(v)}$ SU(2) representations. One can define the total quantum number $j_v$ associated to $e(v)$ by the standard sum of SU(2) irreducible representations via SU(2) Clebsch-Gordan coefficients. 

The functions $I_v$ can be expanded in terms of spherical harmonics developed from $\chi_i$ variables, $Y_l^m(\chi)$, {\it i.e.} 
\begin{equation}
I_v(\chi)=\sum_{l_v n_v} d_{l_v n_v}(\chi^2) Y_{l_v}^{n_v}(\chi),\label{iv}
\end{equation}

$d_{l_v m_v}$ being the coefficients of such an expansion, which in general depend on $\chi^2$. Hence, each term inside the sum (\ref{iv}) transforms under rotations according with the irreducible SU(2) representation having $l_v$ as quantum number. Therefore, at the end irreducible representations of the rotation group are determined by summing via SU(2) Clebsch-Gordan coefficients the representations $\{j_{v},l_v\}$, {\it i.e.}
\begin{equation}
|I,M>_v=\sum_{m_v n_v} C^{IM}_{j_{e(v)} l_v m_{e(v)} n_v} |j_{e(v)},{m_{e(v)}}>\otimes Y_{l_v}^{n_v}(\chi).
\end{equation}

The so envisaged representation is particularly useful, because it allows us to define the rotation invariant states simply by projecting down $\psi$ to the trivial representation at each vertex $|0,0>$. The resulting picture resembles that one of invariant states for LQG, where the intertwiners maps the sum of SU(2) representations incoming in each vertex to the trivial one, but here the orbital angular momentum associated with $I_v$ is an additional element of such a sum.

Hence, it is possible to define a rotation-invariant state by summing the representations associated to each edge and the orbital angular momentum of the functional $I_v(\chi)$ and projecting down to the fundamental representation. The LQG states constitute a particular case, in which $I_v(\chi)$ is projected down to the representation with vanishing angular momentum.  

\subsection{The boost generator}

The expression of the boost generator (\ref{boo}) is not uniquely determined on a quantum level, because the piece $\chi_i\chi_j\pi^j$ is ambiguous. We choose the ordering with all momenta on the right and this choice gives a non-symmetric boost operator. 

The generator is made of two terms acting at vertices only. The first one contains the conjugate momenta to $\chi_i$ and it acts on the $\chi$-dependent part only. 

For instance, let us consider the third component of the boost generator. One finds at each vertex
\begin{equation} 
(\pi^3+\chi_3\chi_i\pi^i) (d_{l n}(\chi^2)Y_{l}^{n})=-i( -\chi^3 d'_{l n}Y_{l}^{n}+d_{l n}\partial_3 Y_{l}^{n}),
\end{equation}

where $d'_{ln}=2(1-\chi^2)\partial_{\chi^2} d_{l n}-\frac{1}{2}d_{l n}$ and we dropped the indexes labeling the vertex. Using well-known relations for the derivatives and the product of spherical harmonics (\ref{arm1}), (\ref{arm2}), one gets
\begin{eqnarray}
(\pi^3+\chi_3\chi_i\pi^i) d_{ln}(\chi^2)Y_{l}^{n}=i(d'_{l n}-l d_{ln})\sqrt{\frac{(l+1)^2-n^2}{(2l+1)(2l+3)}}Y_{l+1}^{n}+\nonumber\\+i(d'_{ln}+(l+1) d_{l n})\sqrt{\frac{l^2-n^2}{(2l-1)(2l+1)}}Y_{l-1}^{n}.\label{form}
\end{eqnarray}

The second term in (\ref{form}) is made of the SU(2) generators times $\chi_i$, thus it involves both the $\chi$-dependent functions and the SU(2) degrees of freedom. In particular, it reads as
\begin{eqnarray}
\beta\epsilon_{3}^{\phantom{1}ij}\chi_iG_j |j,m>\otimes d_{ln}Y_{l}^{n}=\frac{i}{2}\beta d_{ln}(\chi_+G_--\chi_-G_+)|j,m>\otimes Y_{l}^{n}=\nonumber\\=-\frac{i}{2}\beta\chi \frac{d_{ln}}{\sqrt{2l+1}}\bigg(\sqrt{(j+m)(j-m+1)}\sqrt{(l+n+2)(l+n+1)}|j,m-1>\otimes Y^{n+1}_{l+1}-\nonumber\\-
\sqrt{(j+m)(j-m+1)}\sqrt{(l-n-1)(l-n)}|j,m-1>\otimes Y^{n+1}_{l-1}-\nonumber\\-\sqrt{(j-m)(j+m+1)}\sqrt{(l-n+2)(l-n+1)}|j,m+1>\otimes Y^{n-1}_{l+1} 
+\nonumber\\+\sqrt{(j-m)(j+m+1)}\sqrt{(l+n-1)(l+n)}|j,m+1>\otimes Y^{n-1}_{l-1}\bigg).  
\end{eqnarray}

By summing the two expressions above, one finds the final expression for the third components of the boost generator:
\begin{eqnarray}
K_3|j,m>\otimes d_{ln}Y_{l}^{n}=i(d'_{l n}-l d_{ln})\sqrt{\frac{(l+1)^2-n^2}{(2l+1)(2l+3)}}|j,m>\otimes Y^{l+1}_{n}+\nonumber\\+i(d'_{ln}+(l+1) d_{l n})\sqrt{\frac{l^2-n^2}{(2l-1)(2l+1)}}|j,m>\otimes Y^{l-1}_{n}-\nonumber\\-\frac{i}{2}\beta\chi \frac{d_{ln}}{\sqrt{2l+1}}\bigg(\sqrt{(j+m)(j-m+1)}\sqrt{(l+n+2)(l+n+1)}|j,m-1>\otimes Y^{n+1}_{l+1}-\nonumber\\-
\sqrt{(j+m)(j-m+1)}\sqrt{(l-n-1)(l-n)}|j,m-1>\otimes Y^{n+1}_{l-1}-\nonumber\\-\sqrt{(j-m)(j+m+1)}\sqrt{(l-n+2)(l-n+1)}|j,m+1>\otimes Y^{n-1}_{l+1} 
+\nonumber\\+\sqrt{(j-m)(j+m+1)}\sqrt{(l+n-1)(l+n)}|j,m+1>\otimes Y^{n-1}_{l-1}\bigg).
\end{eqnarray}

The action of other components can be evaluated following the same procedure. 

As soon as LQG states are concerned, the action of the first term inside the boost operator vanishes in the adopted operator ordering for construction (\ref{pi}), while the second term does not provide any contribution because of SU(2) gauge invariance. This can be verified for the third component using equations above by writing $I^{LQG}=d^{LQG}_{00}Y^0_0$ with $(d^{LQG}_{00})'=0$. 

Therefore, LQG states are invariant under boost transformations. 

\section{Conclusions} 

In this work it has been demonstrated how it is possible to reduce the configuration space proper of the BF theory to the kinematical Hilbert space of LQG in the framework of LQG without the time gauge. The key points of this reduction have been i) the projection of Lorentz irreducible representations to the ones associated with the Ashtekar-Barbero connections (\ref{rep}), which provides the proper solution of second-class constraints, ii) the restriction to SU(2) invariant intertwiners, which is a consequence of the imposition of Hamiltonian constraints on kinematical states. This analysis can be considered as the quantum counterpart of LQG without the time gauge \cite{prl} and it outlines the usefulness of such a formulation for gravity in a covariant setting. 

In particular, within this scheme it is possible to investigate the implications of the restriction to a certain subset of Lorentz irreps (associated with Ashtekar-Barbero connections). We outlined how once such a restriction is made the action of the operators associated with reduced phase-space coordinates coincides with the one predicted solving the constraints classically. This way, it has been possible to address the point raised in \cite{alsf} on the consistency between the implementation of constraints on a quantum level and a reduced phase space quantization.   

Furthermore, an alternative procedure to infer the kinematical Hilbert space of LQG from the configuration space of BF theories has been provided. The kinematical Hilbert space is here endowed with a basis given by SU(2) spin-networks and the same Hilbert space structure as the one of LQG is defined without fixing $\chi_i$. We analyze the behavior of rotations and boosts in this framework and we found that LQG states are invariant under both kind of transformations. 

The resulting scenario differs from the one envisaged in the context of projected spin-networks \cite{psn}, where the projection to the $SU_\chi(2)$ subgroup takes place at vertices only. In this respect, the implementation of this scheme in a discrete setting will clarify the differences between these scenarios. In particular, the dynamical features must be investigated via the analysis of the path-integral formulation in the adopted set of phase-space coordinates along the lines of \cite{sft}. Such an analysis will determine the relevance of the proposed approach for the Quantum Gravity issues. This work can also be considered as the first step in this direction, because it establishes a clear correspondence between LQG and BF on the boundary hypersurfaces of a path integral formulation. What remain to be done is essentially the description of a discretized space-time in terms of the adopted set of variables. BF models being topological, the whole dynamical information is contained in the Clebsch-Gordan coefficients of the Lorentz group. Henceforth, the investigation on the dynamical correspondence with LQG will require the analysis on the properties of the Clebsch-Gordan coefficients of the Lorentz group as soon as the restriction to SU(2) representations take place. This study may also shed light on the parameter $r$, which is ambiguity in the proposed scheme. Moreover, the role of the cross-simplicity constraints arising in a discretized framework need to be clarified in order to make a real comparison with existing Spin-Foam models.       

\appendix

\section{Relations for harmonic functions}

Recurrence relations for associated Legendre functions $P^m_l(\mu)$: 
\begin{eqnarray}
\mu P^m_l(\mu)=\frac{l-m+1}{2l+1}P^m_{l+1}(\mu)+\frac{l+m}{2l+1}P^m_{l-1}(\mu),\\
(\mu^2-1)\partial_\mu P_l^m(\mu)=\frac{l(l-m+1)}{2l+1}P^m_{l+1}(\mu)-\frac{l+1(l+m)}{2l+1}P^m_{l-1}(\mu).
\end{eqnarray}

Recurrence relations for harmonic functions $Y_l^m(\chi)=(-1)^m \frac{\sqrt{2l+1}}{\sqrt{4\pi}}\sqrt{\frac{(l-m)!}{(l+m)!}}P^m_l(cos(\theta_\chi))$:
\begin{equation}
\chi_3 Y_l^m(\chi)=\chi \left(\sqrt{\frac{l^2-m^2}{(2l+1)(2l-1)}}Y^m_{l-1}+\sqrt{\frac{(l+1)^2-m^2}{(2l+3)(2l+1)}}Y^m_{l+1} \right).\label{arm1}
\end{equation}

\begin{equation}
\frac{\partial}{\partial \chi_3}Y^m_l(\chi)=\frac{1}{\chi}\left(-(l+1)\sqrt{\frac{l^2-m^2}{(2l+1)(2l-1)}}Y^m_{l-1}+l\sqrt{\frac{(l+1)^2-m^2}{(2l+3)(2l+1)}}Y^m_{l+1} \right).\label{arm2}
\end{equation}

\end{document}